# Involution and BSConv Multi-Depth Distillation Network for Lightweight Image Super-Resolution

Akram Khatami Rizi
Cyberspace Research Institute, Shahid Beheshti University
Tehran Iran
a.khatamirizi@mail.sbu.ac.ir

Ahmad Mahmoudi-Aznaveh
Cyberspace Research Institute, Shahid Beheshti University
Tehran Iran
a_mahmoudi@sbu.ac.ir

***Abstract*—Single-image super-resolution (SISR) is a fundamental problem in computer vision that aims to reconstruct high-resolution (HR) images from low-resolution (LR) inputs. Although convolutional neural networks (CNNs) have achieved substantial advancements, deeper architectures often introduce excessive parameters, higher memory usage, and computational cost, limiting their applicability on resource-constrained devices. Recent research has thus focused on lightweight architectures that preserve accuracy while reducing complexity. This paper presents the Involution and BSConv Multi-Depth Distillation Network (IBMDN), a lightweight and effective architecture for SISR. The proposed IBMDN comprises Involution and BSConv Multi-Depth Distillation Blocks (IBMDB) and a Contrast and High-Frequency Attention Block (CHFAB). IBMDB employs varying combinations of Involution and BSConv at multiple depths to perform efficient feature extraction while minimizing computational complexity. CHFAB, a lightweight self-attention mechanism, focuses on extracting high-frequency and contrast information to enhance perceptual quality in the reconstructed images. The flexible design of IBMDB enables it to be seamlessly integrated into diverse SISR frameworks, including information distillation, transformer-based, and GAN-based models. Extensive experiments demonstrate that incorporating IBMDB significantly reduces memory usage, parameters, and floating-point operations (FLOPs), while achieving improvements in both pixel-wise accuracy and visual quality. The source code is available at: https://github.com/akramkhatami/IBMDN.**

## I. INTRODUCTION

With the rapid growth of social networks and the expansion of image content exchange, high-resolution (HR) image recovery from low-resolution (LR) inputs, known as single-image super-resolution (SISR), has become a fundamental challenge in computer vision. Recent advances in deep learning have led to numerous CNN-based models [1-9], generally classified as GAN-based or non-GAN-based. While deeper networks often improve reconstruction quality, they also increase memory usage, parameter count, and computational cost, limiting their applicability in resource-constrained environments. To address these issues, lightweight and efficient architectures have been proposed, leveraging techniques such as recurrent structures [4], group convolutions (GConv) [10], information distillation [11-15], self-attention mechanisms[12, 14, 16, 17], and network architecture search (NAS) [18]. Additionally, alternative operators have been introduced to reduce redundancy in standard convolution and improve feature extraction at a lower computational cost [19-22].

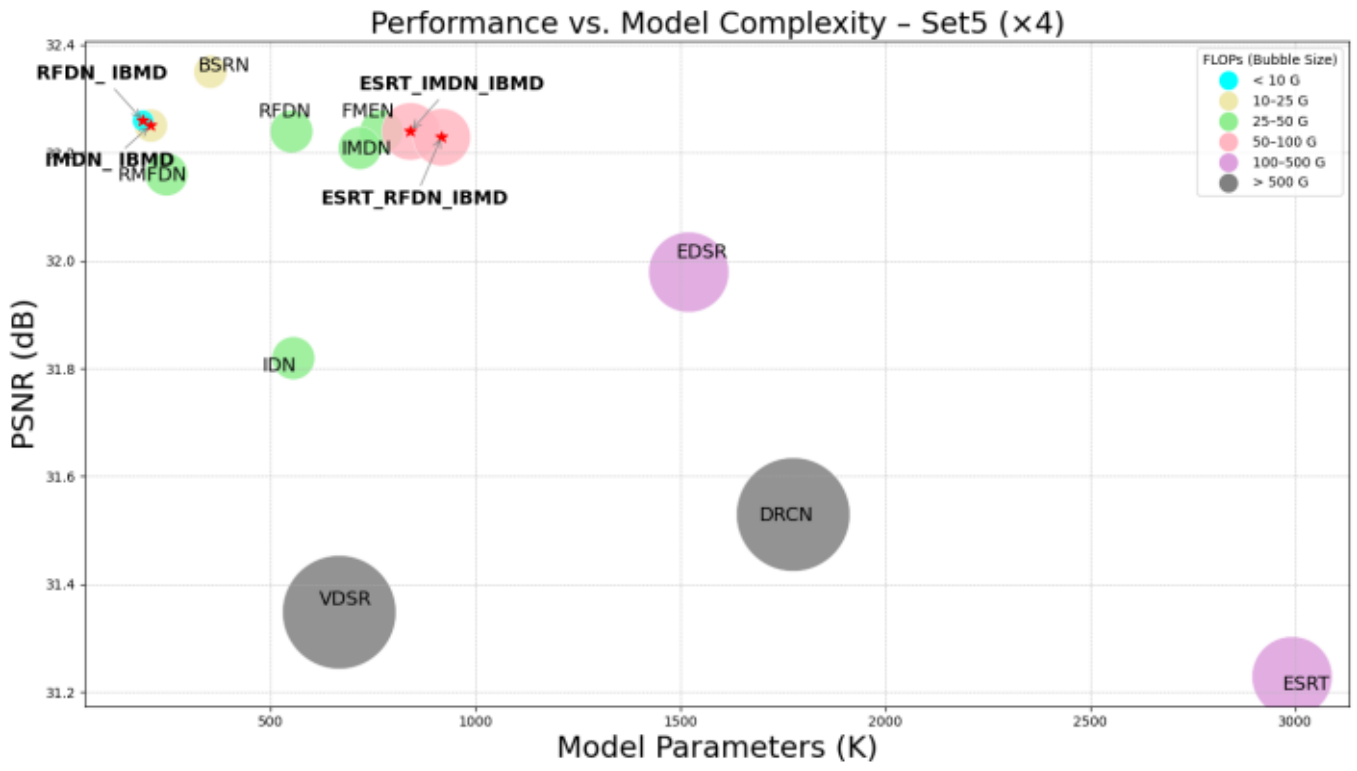

*Figure 1. Performance and model complexity comparison on the Set5 dataset for upscaling factor ×4.*

Among existing techniques, information distillation and self-attention have proven effective for balancing efficiency and performance in SISR. Motivated by this, we propose the Involution and BSConv Multi-Depth Distillation Network (IBMDN), a lightweight and efficient architecture that improves reconstruction quality in two ways. First, Involution and BSConv Multi-Depth Distillation Blocks (IBMDB) use depth-dependent structures that combine pixel-level (BSConv) and visual-based (Involution) operators, enabling simultaneous extraction of pixel and visual features. Unlike conventional SISR models with uniform deep feature extraction layers, IBMDB adapts block structures to different depths, achieving efficient and high-quality super-resolution. Second, the Contrast and High-Frequency Attention Block (CHFAB) enhances the network's ability to extract contrast and high-frequency information, improving sharpness and perceptual quality. As shown in Fig. 1, IBMDN delivers competitive results compared to some existing SISR methods while requiring fewer parameters, lower computational cost, and faster inference.

The main contributions of this paper are summarized as follows:

- Proposing IBMDN for high-resolution image reconstruction, which employs Involution and BSConv Multi-Depth Distillation Blocks (IBMDB) and the Contrast and High-Frequency Attention (CHFAB) module, achieving outstanding results with minimal parameters and FLOPS.

- IBMDB's feature extraction blocks employ a multi-depth information distillation mechanism that adaptively combines pixel-wise and perceptual operators at each depth, improving accuracy while significantly reducing computational complexity.
- CHFAB, a lightweight self-attention module, enhances fine-detail reconstruction by extracting high-frequency information and capturing both local and global interactions.
- The proposed IBMDB deep feature extraction block achieves strong performance in information distillation-based architectures with minimal parameters and FLOPs. Experiments show that integrating it into transformer- and GAN-based models improves evaluation metrics while significantly reducing computational cost.

## II. RELATED WORK

### A. Single Image Super-Resolution Based on CNN

In recent years, Single Image Super-Resolution (SISR) has received growing attention in computer vision research, reflecting its crucial role in visual quality enhancement. With the rise of deep learning, Convolutional Neural Networks (CNNs) have become the dominant approach for SISR. SRCNN [1], as the first CNN-based model, employed a simple three-layer design, while later models, such as VDSR [2] and FSRCNN [5] improved reconstruction accuracy with deeper architectures. Shi et al. [23] further accelerated inference through a sub-pixel convolution layer. Although deeper networks enhance accuracy, they increase computational cost. To address this, recursive networks reused a fixed set of parameters across layers [4]. Residual networks, by learning the difference between low-resolution (LR) and high-resolution (HR) images, enhance high-frequency detail reconstruction [3, 7]. Additionally, Kim et al. [3] introduced a global residual connection between the first and last layers to mitigate vanishing gradients and facilitate LR information flow into deeper layers.

With the success of transformer architectures in natural language processing, combining transformers with convolutional operations for SISR has gained attention [9, 24]. Transformers capture complex relationships and extract richer image features, crucial for accurate reconstruction, while convolutional networks focus on local details, ensuring higher image quality. Integrating both approaches enhances reconstruction, although transformer architectures typically involve a large number of parameters and high computational cost.

Generative Adversarial Networks (GANs) have been used to improve perceptual quality [7, 8, 25]. A GAN consists of a generator that creates realistic textures and a discriminator that evaluates them against real images. Although GANs perform well in enhancing perceptual quality, particularly for textures, they face challenges in accurately reconstructing edges and preserving structural details. Their application is further limited by the large number of parameters and high computational complexity.

### B. Lightweight Single Image Super-Resolution

Lightweight SISR models aim to optimize performance in resource-constrained systems by reducing computational complexity through fewer parameters, efficient filters, and techniques like quantization and pruning. However, such reductions can affect reconstruction quality, so recent research focuses on models that maintain or improve image quality while minimizing computational costs. To address these challenges, lightweight SISR approaches include changes in neural network architectures and the use of neural architecture search (NAS). Changes in architecture include techniques such as parameter sharing, recursive architectures, optimal block combinations, group convolution, information distillation, and attention mechanisms. Overall, these approaches aim to balance efficiency and performance in lightweight SISR models.

One of the earliest lightweight SISR models was FSRCNN, which introduced sampling at the end of the network [5]. Subsequent models, such as CARN [26] and MADNet [27], improved efficiency by reducing parameters, although accuracy remained limited for some applications. Recursive models like DRCN enhanced processing speed through parameter sharing across layers [4]. Group convolution, first proposed in AlexNet [10] and later optimized in ShuffleNet [19, 28], increased efficiency via channel splitting and parallel processing. Block-design strategies in LatticeNet [29] and MAFFSRN [30] showed that optimal block combinations can improve performance. ESRT [9] leveraged high-frequency filters and channel splitting to achieve strong results. These techniques are particularly suitable for devices with limited computational resources. Despite these advancements, some lightweight models still fall short in terms of resource efficiency and reconstruction accuracy, facing real-world challenges such as high memory consumption, FLOPs, and computational complexity. Designing truly lightweight, fast, and accurate SISR models thus remains an open challenge.

**Optimization of Neural Network Operators.** Convolution is a core operation in CNNs, vital for detecting local patterns but computationally expensive due to numerous filters and complex operations. Depthwise Separable Convolution (DSConv) addresses this challenge by splitting the process into two stages [19, 20]. In the first stage, depthwise convolution applies a single filter to each input channel separately, significantly reducing FLOPs. In the second stage, pointwise convolution merges inter-channel information. While this approach lowers the number of parameters and accelerates processing, the strictly sequential structure can lead to partial loss of feature information. To alleviate this, BSConv reverses the DSConv order and exploits correlations between kernels, enabling richer feature extraction. This reordering improves accuracy without adding computational overhead, making BSConv a compelling alternative for lightweight models [21].

Involution, a more recent innovation, takes a different approach by focusing on spatial adaptivity [22]. Unlike standard convolution, it generates dynamic, location-specific kernels for each pixel, making it more sensitive to spatial variations and capable of modeling long-range dependencies. While conventional convolution limits cross-channel interactions, Involution enables richer channel information exchange while reducing redundant computation. As shown in Figure 3(b), a group filter is first applied to each pixel along the channel dimension. The resulting output is then reshaped into a new dimension matrix and used as a local feature-based filter. This design achieves a better trade-off between representational power and efficiency, minimizing both parameters and FLOPs, which makes it suitable for deployment in hardware-limited scenarios.

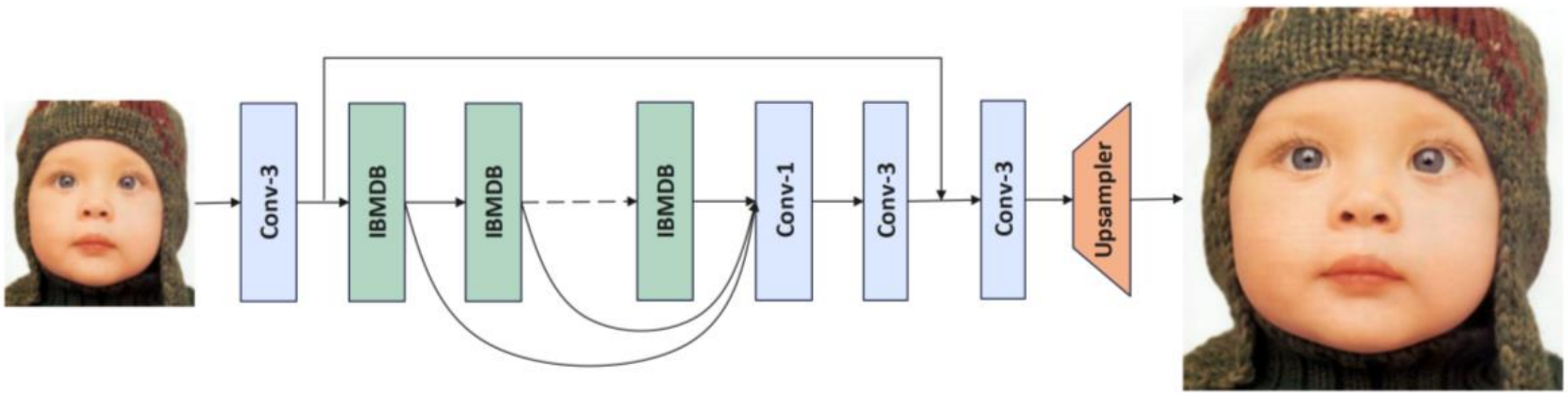

*Figure 2. The architecture of the Involution and BSConv Multi-Depth Distillation Network (IBMDN).*

**Information Distillation.** Information distillation is a technique for creating lightweight models by using channel splitting to extract features, first proposed by Hui et al. with the Information Distillation Network (IDN) [11]. In IDN, extracted features are divided into two parts: one part is retained, while the other is further processed. This approach reduces computational complexity, optimizes memory usage, and lowers resource consumption. Building on IDN, Hui introduced the IMDN model, which incorporates the CCA and IMDB blocks [12]. IMDN achieves faster and more accurate performance compared to IDN but still has a relatively large size, with *700K* parameters. In many subsequent studies, the IMDB block has served as a foundational building block for information distillation, and modifications to its structure have led to new lightweight models with improved PSNR and SSIM metrics.

Despite its advantages, channel splitting can limit model flexibility. To address this, the Residual Feature Distillation Block (RFDB) was proposed as a more flexible and lightweight alternative to IMDB [13]. RFDB employs a Shallow Residual Block (SRB) to preserve spatial information and enhance feature extraction. It consists of a $3\times3$ convolution, an identity connection, and an activation unit, optimizing residual learning without adding complexity. Additionally, the identity branch features are refined with a $1\times1$ convolution, further increasing flexibility. Compared to IMDB, RFDB is lighter, more efficient, and more adaptable, though fine-tuning remains necessary for optimal performance. Building on RFDB, the BSRN model introduced blueprint separable convolution (BSConv) to reduce redundancy and enhance performance, while incorporating an efficient self-attention module (ESA) alongside CCA to improve attention capabilities [14]. Finally, the RMFDN model integrates Involution into the core structure of RFDN, achieving comparable results while further reducing the number of parameters [15].

**Attention Mechanism in Super-Resolution**. Recently, attention mechanisms have become increasingly popular in Single Image Super-Resolution (SISR) systems. Attention blocks are commonly categorized into Channel Attention (CA) [12, 31], Spatial Attention (SA) [14, 32], Pixel Attention (PA) [17], and others. However, most existing attention modules focus primarily on local interactions and often incur high memory costs. Incorporating both global and local interactions can more effectively enhance high-frequency information recovery.

To overcome these limitations, Zongcai Du introduced the High-Frequency Attention Block (HFAB), a novel attention module designed to enhance high-frequency information in SR tasks [16]. Its main goal is to detect edges through a learning-based process. While combining neighboring pixels helps edge detection, standard convolutions are limited, capturing mostly local dependencies when evaluating each pixel. To overcome this, batch normalization (BN) is applied to incorporate global interactions and improve the effectiveness of the sigmoid activation. Skip connections are also included to further optimize feature extraction and preserve important details.

## III. PROPOSED METHOD

### A. Framework

In this section, we detail the Involution and BSConv Multi-Depth Distillation Network (IBMDN). The overall architecture, shown in Figure 2, consists of four key parts: shallow feature extraction, deep feature extraction, feature fusion, and reconstruction, which together form the complete network for SISR. The input image ($I_{LR}$) is initially mapped into the feature space $F_0$ by the shallow feature extraction module.

$$F_0 = H_{SF}(I_{LR}) \quad (1)$$

In this step, the shallow feature extraction module $H_{SF}$ uses a $3\times3$ convolution to extract the initial features. The output $F_0$ is subsequently fed into multiple IBMDB blocks for deep feature extraction, expressed as follows:

$$F_k = H_k(F_{k-1}) \; k = 1,\ldots,n \quad (2)$$

In this relation, $H_k$ represents the $k$-$th$ block, while $F_{k-1}$ and $F_k$ denote its input and output features, respectively. The features extracted at different levels are first combined using a $1\times1$ convolution and then refined with a $3\times3$ convolution. Accordingly, the feature fusion can be formulated as follows:

$$F_{fused} = H_{fusion}(Concat(F_1, \ldots, F_k)) \quad (3)$$

Here, $H_{fusion}$ denotes the fusion section and $F_{fused}$ the aggregated features. A global skip connection is used to leverage residual learning, and the reconstruction is formulated as follows:

$$I_{SR} = H_{rec}(F_{fused} + F_0) \quad (4)$$

In this step, $H_{rec}$ includes a $3\times3$ convolution layer and a pixelshuffle operation as the reconstruction module.

### B. *Involution and BSConv Multi-Depth Distillation Block*

The Involution and BSConv Multi-Depth Distillation Block (IBMDB) comprises three Involution-BSConv Residual Blocks (IBRBs) and a Contrast and High-Frequency Attention Block (CHFAB), as shown in Figure 3(a). Standard *3×3* convolutions are not always optimal for lightweight super-resolution networks due to their high computational cost and large parameter count [33, 34]. To overcome these limitations, BSConv replaces standard convolutions, enhancing intra-kernel correlation, effectively reducing the number of parameters while better preserving information and improving model accuracy without increasing architectural complexity. In addition, Involution is incorporated into the residual modules. With its self-attention mechanism and low parameter count, Involution excels at capturing visual features. Therefore, a hybrid approach using BSConv and Involution is adopted to improve both efficiency and feature extraction quality.

In SISR architectures, feature extraction blocks are typically constructed with a uniform structure, regardless of the network's depth. However, features at different layers vary, and a fixed block structure can limit the network's capacity to utilize these differences. Adapting the block structure according to the network's depth can therefore enhance performance. Super-resolution mainly relies on convolutional operations emphasizing pixel-level fidelity, so metrics like PSNR and SSIM are commonly used, while visual quality is often overlooked. By designing a network that simultaneously considers both pixel-level and perceptual metrics, it is possible to improve performance in both aspects. For this purpose, the proposed architecture adapts block structures at different depths to balance pixel-focused and perceptual feature learning.

Building on insights from prior works [33, 34], the proposed feature extraction blocks integrate both pixel- and visual-focused operators in a depth-aware manner. Pixel operators are most effective in the early layers, where feature maps have undergone minimal changes and local pixel similarity is still high. In these layers, BSConv is preferred due to its efficiency and strong pixel-level correlation with minimal parameter overhead. As the network deepens, extracting high-level visual features becomes increasingly important. This requires mechanisms that model inter-channel dependencies and abstract representations. In this context, Involution acts as a self-attention-based operator capable of capturing inter-channel dependencies and high-level features.

Thus, in early layers, BSConv maintains pixel fidelity and local interactions, while in deeper layers, Involution dominates for capturing global and semantic information. Denoting BSConv as $B_S$ and Involution as $I$, the optimal operator sequence for deep feature extraction, based on previous empirical studies [33], is: $B_SB_SB_S – B_SB_SB_S – B_SIB_S – B_SIB_S – IB_SI – III$.

$$
\begin{aligned}
&F^n_{refined_1},\ F^n_{coarse_1} = Split^n_1(CL^n_1(F^n_{in})),\\
&F^n_{refined_2},\ F^n_{coarse_2} = Split^n_2(CL^n_2(F^n_{coarse_1})),\\
&F^n_{refined_3},\ F^n_{coarse_3} = Split^n_3(CL^n_3(F^n_{coarse_2})),\\
&F^n_{refined_4} = CL^n_4(F^n_{coarse_3})
\end{aligned} \tag{5}
$$

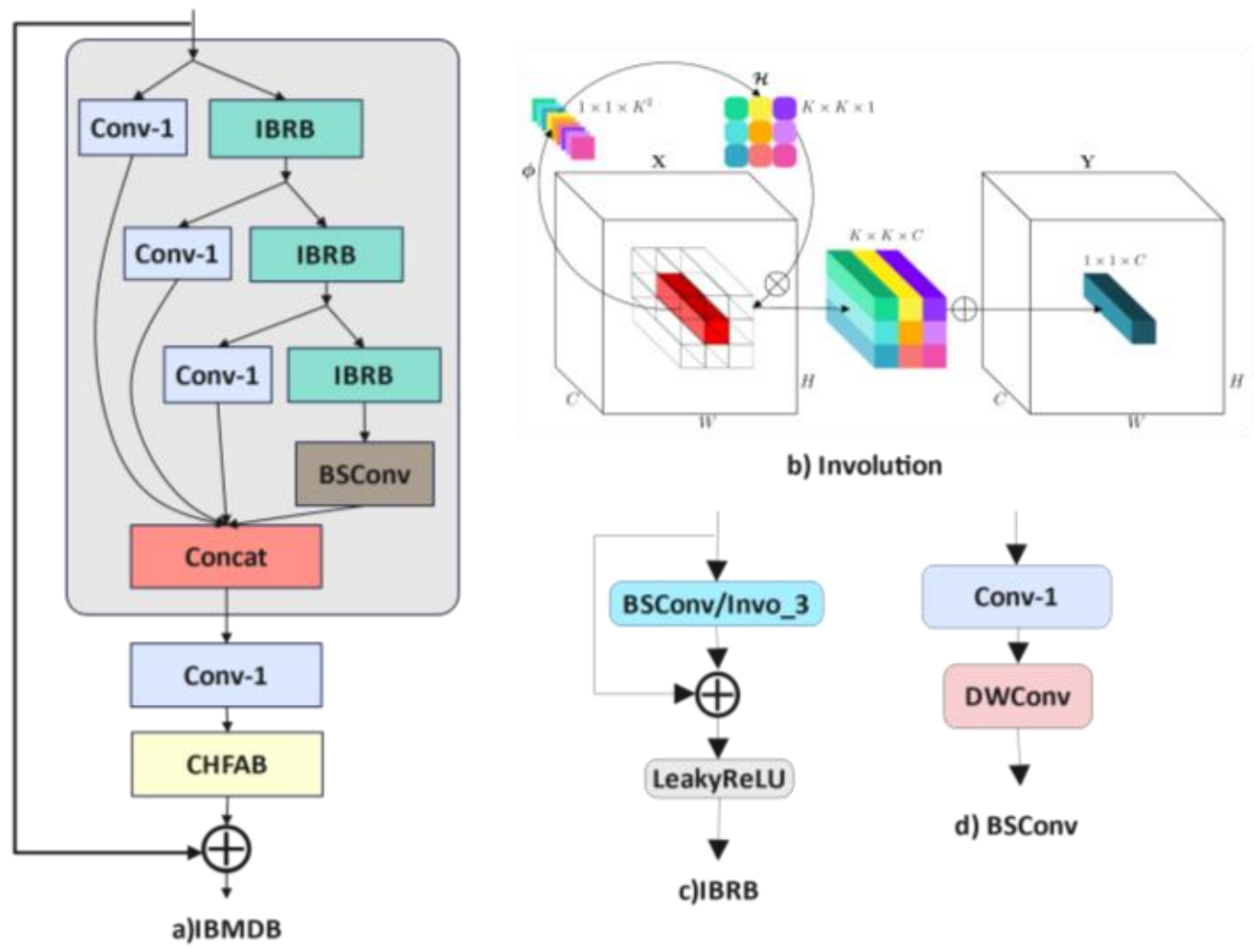

*Figure 3. (a) The architecture of the proposed IBMDN. (b) The architecture Involution [22]. (c) The architecture of the proposed IBRB in IBMDN. (d) The architecture of BSConv [21].*

where $CL^n_j$ denotes the *j-th* BSConv or involution layer of the *n-th* IBMDB, $Split^n_j$ denotes the *j-th* channel splitting layer of the *n-th* IBMDB, $F^n_{refined\text{-}j}$ represents the *j-th* refined features, and $F^n_{coarse\text{-}j}$ is the *j-th* coarse features to be further processed. Feature maps are then concatenated along the channel dimension.

$$
\begin{aligned}
&F^n_{distilled} =\\
&Concat(F^n_{refined_1}, F^n_{refined_2}, F^n_{refined_3}, F^n_{refined_4})
\end{aligned} \tag{6}
$$

where *Concat* denotes the concatenation operation along the channel dimension.

### C. *Contrast and High-Frequency Attention Block*

Attention mechanisms have become a cornerstone in image super-resolution (SR) architectures. Despite the additional memory and parameter overhead associated with their structural complexity, their substantial contribution to enhancing network performance justifies their widespread adoption. Most attention blocks concentrate on local interactions, often overlooking the potential benefits of integrating global and local dependencies, which can significantly enhance high-frequency information recovery. Previously, Zongcai Du introduced the High-Frequency Attention Block (HFAB) to address these limitations, employing a residual structure and batch normalization to enhance both local and global feature interactions [16].

In this study, we adopt HFAB as the base attention block and introduce a lightweight variant, the Contrast and High-Frequency Attention Block (CHFAB). As illustrated in Figure 4, CHFAB integrates contrast information and mean aggregation within its structure to further enhance visual quality. Furthermore, to minimize parameter count, Involution, an efficient self-attention-based visual operator, is integrated with BSConv within the attention block.

## IV. EXPERIMENTS

### A. *Datasets and Metrics*

Following standard practice in the literature, we use the DIV2K dataset, comprising *800* high-resolution training images, for training in image super-resolution tasks. For evaluation, four well-established benchmark datasets are used: Set5, Set14, BSD100, and Urban100. The quality of the super-resolved images is evaluated using the metrics of PSNR, SSIM, and Learned Perceptual Image Patch Similarity (LPIPS) [35]. All evaluations are conducted on the Y channel of the YCbCr color space. Mean Absolute Error (MAE) combined with perceptual loss [36] is adopted as the loss function to quantify the differences between the super-resolved images and the ground truth.

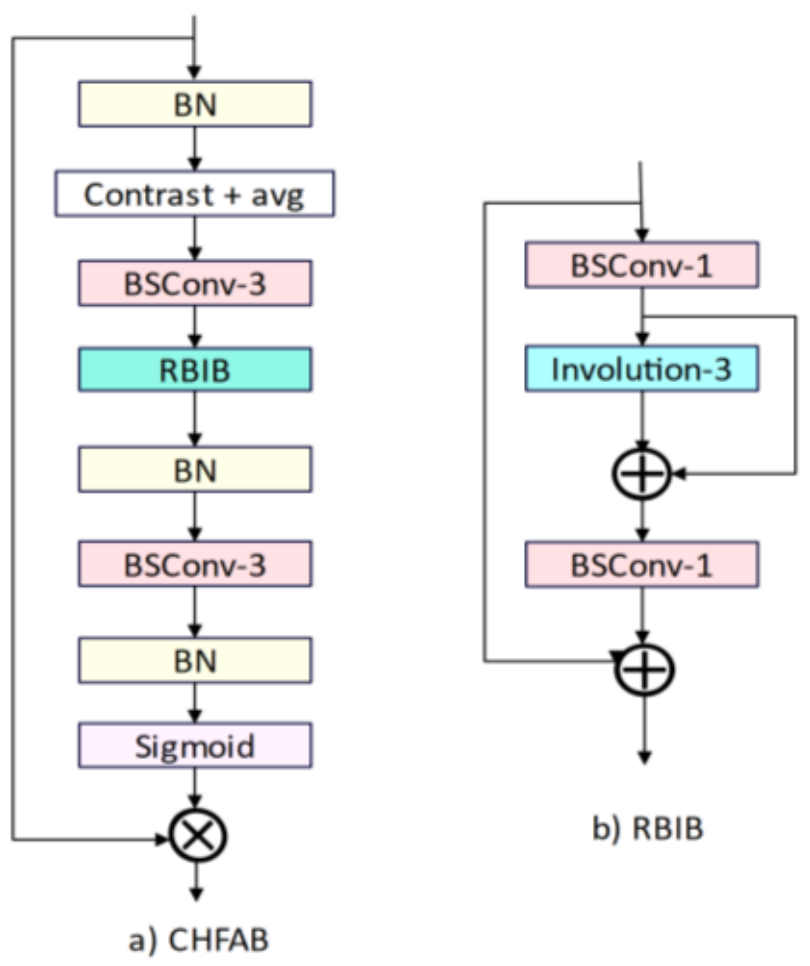

*Figure 4. The Contrast and High-Frequency Attention Block (CHFAB).*

### *B. Implementation details*

Following the standard information distillation models for LR training images in DIV2K, we downscale HR images using scaling factors of *×2*, *×3*, and *×4*. HR image patches of size *192×192* are randomly cropped from HR images as input to our model, and the mini-batch size is set to *16*. We train our model using the ADAM optimizer with $\beta_1 = 0.9$, $\beta_2 = 0.999$. The initial learning rate is set to $2\times10^{-4}$ and is halved every $2\times10^{5}$ iterations. We set the number of IBMDB to *6* in our proposed model. The network is implemented using the PyTorch framework and trained on Google Colab Pro.

### *C. Model analysis*

**The Impact of Involution on Network Depth in Super-Resolution Models.** In SISR, both the visual quality and the pixel-level accuracy of the image influence the performance of the network. In our previous research, we investigated the impact of involution on network depth using seven different IMDN models [33]. In these models, as the network becomes deeper, the use of involution compared to convolution increases, and the block structures vary accordingly. To represent these changes, the combination of convolution *(C)* and involution *(I)* is expressed as *CCC-CCC-CIC-CIC-ICI-III*, which we define as the Involution and Convolution Multi-Depth (ICMD) structure.

For effective evaluation, we replaced the deep feature extraction component with ICMD in the RFDN and BSRN models. As presented in Table 1, the RFDN model with four blocks (*CCC-CIC-ICI-III*), following the original design, showed a slight performance drop compared to the baseline. Given that the ICMD structure is more suitable for deeper networks, we also evaluated a six-block RFDN with the complete ICMD configuration (*CCC-CCC-CIC-CIC-ICI-III*). For model comparison, we considered only the PSNR and SSIM metrics. The baseline models were retrained, and all experiments were conducted over *100* epochs.

According to the results, our proposed method achieved comparable or superior performance to the baseline IMDN, RFDN, and BSRN models in fewer training epochs and with fewer parameters. The combined use of pixel-level and visual-oriented operators improved both pixel-based and perceptual metrics, often conflicting in GAN-based models. As the IMDB and RFDB blocks constitute the fundamental components of information distillation, they will be employed in our subsequent evaluations of the IMDN_ICMD and RFDN_ICMD models with six blocks.

*TABLE I. The Impact of Involution on Network Depth in Super-Resolution Models in the datasets Set5, Set14, BSD100, and Urban100. scale ×2 and with 100 epochs.*

| Model | Params | Set5 | Set14 | BSD100 | Urban100 |
|---|---|---|---|---|---|
| | | PSNR/SSIM | | | |
| IMDN | 694K | 37.52/0.9584 | 33.17/0.9119 | 31.79/0.8637 | 31.71/0.9249 |
| RFDN | 534K | 37.55/0.9560 | 33.20/0.9129 | 31.75/0.8961 | 31.69/0.9256 |
| BSRN | 332k | 37.20/0.9552 | 33.68/0.9162 | 31.83/0.8975 | 31.79/0.9281 |
| IMDN_ICMD | 494K | 37.54/0.9575 | 33.18/0.9129 | 31.88/0.8948 | 30.65/0.9207 |
| RFDN_ICMD 4B | 262K | 37.20/0.953 | 32.9/0.9107 | 31.43/0.8905 | 30.55/0.9173 |
| RFDN_ICMD 6B | 424K | 37.63/0.9677 | 33.22/0.9133 | 31.95/0.8950 | 30.71/0.9272 |
| BSRN_ICMD | 340K | 37.42/0.9551 | 33.16/0.9126 | 31.87/0.8946 | 30.58/0.9192 |

**Involution and BSConv Multi-Depth.** BSConv, by reducing parameters and complexity, increases model accuracy; therefore, in the IMDN_ICMD and RFDN_ICMD models, we use the combination of BSConv and Involution instead of standard convolutions. If we denote BSConv as *($B_S$)* and Involution as *(I)*, the distillation block combination in both models will be as follows: $B_SB_SB_S$ - $B_SB_SB_S$ - $B_SIB_S$ - $B_SIB_S$ - $IB_SI$ - *III*. We have named the combination of BSConv and Involution as Involution and BSConv Multi-Depth (IBMD). According to Table 2, the BSConv structure significantly improves the performance of evaluation metrics and reduces the network parameters.

*TABLE II . Performance of the Involution and BSConv combination. The models have been evaluated on the SET5 dataset and ×2 scale with 100 epochs.*

| Model | Parameter | PSNR↑ | SSIM↑ | LPIPS↓ |
|---|---|---|---|---|
| IMDN_ICMD | 494K | 37.54 | 0.9675 | 0.0157 |
| IMDN_ IBMD | 161K | 37.57 | 0.9659 | 0.0150 |
| RFDN_ ICMD | 424K | 37.63 | 0.9677 | 0.0148 |
| RFDN_ IBMD | 150K | 37.67 | 0.9661 | 0.0142 |

**Evaluation of Attention Block Performance.** To evaluate the impact of attention blocks on high-frequency information reconstruction and network performance, a comparative analysis of CCA, ESA, HFAB, and our proposed block (CHFAB) was conducted. For this evaluation, the IMDN_IBMD and RFDN_IBMD models, both of which achieved strong results with fewer parameters, were employed. In these models, four different attention blocks were used in place of the default attention module within the distillation blocks.

As shown in Table 3, while the incorporation of CHFAB slightly increased the parameter count compared to CCA, it achieved markedly better results across all evaluation metrics (PSNR, SSIM, and LPIPS). Moreover, its parameter count remained significantly lower than that of the baseline HFAB, highlighting its superior efficiency and reduced computational complexity.

Since the RFDN_IBMD model offered stronger performance with fewer parameters and greater architectural flexibility, it was chosen as the basis for designing the proposed block. This choice ensures an optimal balance between accuracy, efficiency, and complexity, enabling the final model to excel in reconstructing high-frequency information.

*TABLE III .Evaluation of attention blocks in IMDN_IBMD and RFDN_IBMD models. Implemented on the SET5 dataset and ×2 scale with 100 epochs.*

| ***Model*** | ***Attention block*** | ***Parameter*** | ***PSNR↑*** | ***SSIM↑*** | ***LPIPS↓*** |
|---|---|---|---|---|---|
| IMDN_IBMD | CCA | 161K | 37.57 | 0.9659 | 0.0150 |
| | ESA | 227k | 37.49 | 0.9657 | 0.0163 |
| | HFAB | 394k | 37.34 | 0.9640 | 0.0171 |
| | CHFAB | 185k | 37.71 | 0.9662 | 0.0145 |
| RFDN_IBMD | CCA | 150K | 37.67 | 0.9661 | 0.0142 |
| | ESA | 187k | 37.65 | 0.9647 | 0.0153 |
| | HFAB | 399k | 37.53 | 0.9634 | 0.0155 |
| | CHFAB | 170K | 37.74 | 0.9670 | 0.0142 |

### *D. Lightweight Deep Feature Extraction Structure with IBMDB*

An SISR model typically consists of four main components: shallow feature extraction, deep feature extraction, feature fusion, and reconstruction. In this study, we focus on the deep feature extraction stage. The proposed IBMDB block is lightweight, flexible, and demonstrates strong performance in information distillation models. Several SISR models based on Transformer and GAN architectures have been introduced; however, these architectures are challenging to train due to excessive parameters, computational complexity, and high GPU requirements. For example, the ESRT model at scale ×2, with 64 filters per block, contains nearly 2 million parameters. Similarly, ESRGAN at scale ×4 requires around 43 million parameters. Our proposed deep-style feature extraction structure was employed as the deep extraction component in both the Transformer-based ESRT [9] and the GAN-based ESRGAN [8], showing its broader applicability.

**Transformer Network with IBMDB Structure.** Transformer architectures have demonstrated strong capabilities in modeling long-range dependencies in images, making them increasingly popular in SISR tasks. The Enhanced Super-Resolution Transformer (ESRT) is a hybrid model that combines convolutional layers with Transformer-based modules. In our study, we replaced the original lightweight convolutional block (LCB) in the ESRT structure with our proposed deep feature extraction module, IBMDB. Two variants of the IBMDB structure, IMDN_IBMD and RFDN_IBMD, were independently integrated and evaluated in the ESRT framework.

As presented in Table 4, the ESRT_RFDN_IBMD model achieved superior performance in both PSNR and LPIPS metrics compared to the baseline. To ensure architectural compatibility in the channel divisions in the feature extraction module and the transformer backbone, the number of filters in the IBMDB structure was increased from 50 to 64. The baseline ESRT model was also retrained with the updated configuration for a fair comparison.

TABLE IV. The results of Transformer network with the IBMDB structure were run on the SET5 dataset and ×2 scale with 300 periods.

| ***Model*** | ***Parameter*** | ***PSNR↑*** | ***SSIM↑*** | ***LPIPS↓*** |
|---|---|---|---|---|
| ESRT | 2M | 37.47 | 0.9561 | 0.0187 |
| ESRT_IMDN_ IBMD | 545K | 37.73 | 0.9671 | 0.0146 |
| ESRT_RFDN_ IBMD | 622K | 37.76 | 0.9673 | 0.0144 |

**Generative Adversarial Network with IBMDB Structure.** Recovering texture and fine-grained details remains one of the key challenges in GAN-based super-resolution. Numerous models with GAN architecture have been introduced to address these issues and enhance perceptual quality. ESRGAN, built upon SRResNet, is one such model that improves visual quality by utilizing SRResNet-based generator and discriminator designs. In this study, we incorporate our proposed feature extraction structures, IMDN_IBMD, RFDN_IBMD, ESRT_RFDN_IBMD, and ESRT_IMDN_IBMD, into both the generator and discriminator, replacing the original modules.

As shown in Table 5, all four modified models outperform ESRGAN while using significantly fewer parameters. In many GAN-based super-resolution models, there is often a conflict between pixel-based and perceptual metrics, with PSNR typically being low and thus considered less reliable. However, our approach, which jointly employs BSConv and Involution, improves both types of metrics, demonstrating a more balanced and efficient super-resolution capability.

TABLE V. Results of the Model Generative Adversarial Network with IBMDB structure. Evaluation was performed on the SET5 dataset and the ×4 scale with 700 periods.

| ***Model*** | ***Parameter*** | ***PSNR↑*** | ***SSIM↑*** | ***LPIPS↓*** |
|---|---|---|---|---|
| ESRGAN | G=38.5M<br>D=4.6M<br>43.2M | 21.54 | 0.5227 | 0.1409 |
| IMDN_ IBMD | G= 209K<br>D= 580K<br>787K | 22.62 | 0.5536 | 0.1247 |
| RFDN_ IBMD | G= 657K<br>D= 287K<br>844K | 22.36 | 0.5465 | 0.1310 |
| ESRT_IMDN_IBMD | G= 620K<br>D= 247K<br>867K | 22.99 | 0.5595 | 0.1234 |
| ESRT_RFDN_IBMD | G= 690K<br>D= 323K<br>1M | 22.41 | 0.5514 | 0.1296 |

### *E. Comparison with State-of-the-Art*

The Involution and BSConv Multi-Depth Block (IBMDB), integrated into the IMDN, RFDN, and ESRT architectures, is compared with several state-of-the-art models, including SRCNN [1], VDSR [2], DRCN [4], EDSR [3], FMEN [16], ESRT [24], IMDN [12], RFDN [13], BSRN [14], and RMFDN [15], as shown in Table 6 and Figure 5. Quantitative results for scales ×2, ×3, and ×4 are reported on the Set5, Set14, BSD100, and Urban100 benchmark datasets. Since some models report only PSNR and SSIM, these two metrics are considered for standard evaluation. Models proposed in this study are highlighted in blue.

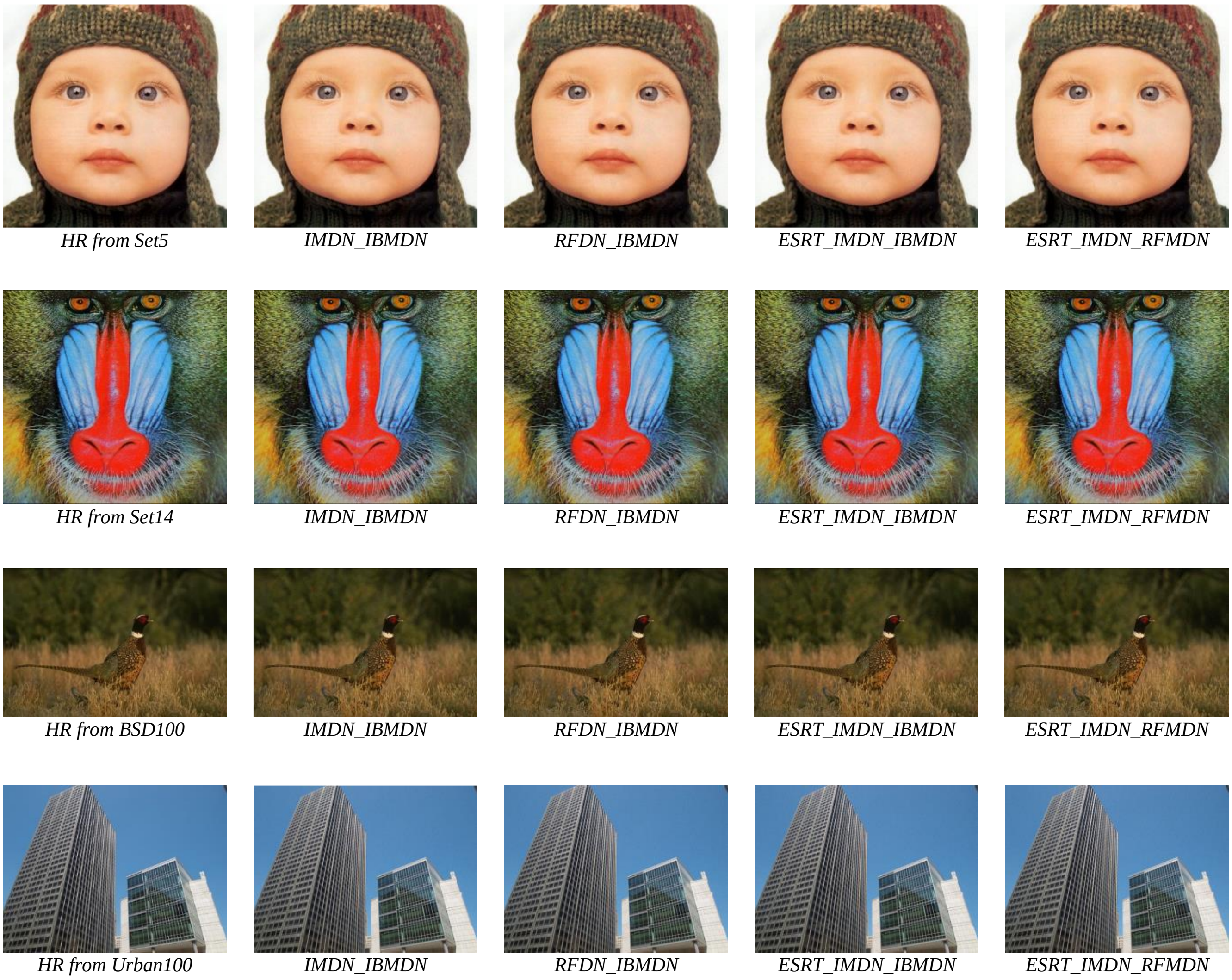

Figure 5. *Visual comparison of IBMDN with information distillation methods in ×4 SR.*

According to the quantitative results in Table 6, our proposed models achieve performance comparable to or better than the baseline and other competitive methods, while maintaining a significantly lower parameter count. Furthermore, a qualitative comparison is provided in Figure 5 using a representative image from the Set5, Set14, BSD100, and Urban100 datasets at ×4 scale. As illustrated, the proposed models are capable of reconstructing high-quality images that are visually close to the ground-truth HR image.

## V. CONCLUSION

In this paper, we introduced the Involution and BSConv-based Multi-Depth Distillation Network (IBMDN) as a lightweight and efficient approach for Single Image Super-Resolution (SISR). The proposed architecture includes Involution and BSConv Multi-Depth Blocks (IBMDB), which optimally combine Involution and BSConv at different depths to strike a balance between reducing computational complexity and maintaining feature extraction quality. Additionally, we employed the Contrast and High-Frequency Attention Block (CHFAB) to enhance high-frequency information and improve visual quality with a low parameter count. The lightweight feature extraction structure of IBMDB was used as the deep feature extraction part in SISR models based on information distillation architectures, transformers, and GAN.

We showed that this structure reduces memory consumption and computational costs while simultaneously increasing pixel accuracy and visual perception of the image. The results show that the proposed method provides efficient performance with low computational cost, and due to the optimal balance between accuracy and efficiency, IBMDN is a suitable choice for resource-constrained devices.

*TABLE VI. Comparison of research results with SISR models. Research results are shown in blue*

| Model | Scal | Params[K] | Multi-Adds[G] | Set5 | Set14 | BSD100 | Urban100 |
|---|---|---|---|---|---|---|---|
| | | | | PSNR/SSIM | | | |
| SRCNN | ×2 | 8 | 52.7 | 36.66/0.5942 | 32.45/0.9067 | 31.36/0.8879 | 29.50/0.8946 |
| VDSR | ×2 | 666 | 612.6 | 37.53/0.9587 | 33.03/0.9124 | 31.90/0.8960 | 30.76/0.9140 |
| DRCN | ×2 | 1774 | 9788.7 | 37.63/0.9588 | 33.04/0.9118 | 31.85/0.8942 | 30.75/0.9133 |
| EDSR | ×2 | 1370 | 316.2 | 37.91/0.9602 | 33.53/0.9172 | 32.15/0.8995 | 31.99/0.9270 |
| FMEN | ×2 | 748 | 172.0 | 38.10/0.9609 | 33.75/0.9192 | 32.26/0.9007 | 32.41/0.9311 |
| ESRT | ×2 | 2697 | 111.47 | 37.47/0.9561 | 33.27/0.9064 | 31.84/0.8975 | 32.03/0.9251 |
| IMDN | ×2 | 694 | 158.8 | 38.00/0.9605 | 33.63/0.9177 | 32.19/0.8996 | 32.17/0.9283 |
| RFDN | ×2 | 534 | 123.0 | 38.05/0.9606 | 33.68/0.9184 | 32.16/0.8994 | 32.12/0.9278 |
| BSRN | ×2 | 332 | 73.0 | 38.10/0.9610 | 33.74/0.9193 | 32.24/0.9006 | 32.34/0.9303 |
| RMFDN | ×2 | 224 | 64.2 | 37.96/0.9605 | 33.54/0.9166 | 32.11/0.8989 | 31.98/0.9268 |
| IMDN_ IBMD | ×2 | 185 | 10.34 | 38.01/0.9662 | 33.65/0.9173 | 32.16/0.8993 | 32.16/0.9279 |
| RFDN_ IBMD | ×2 | 170 | 8.90 | 38.07/0.9670 | 33.68/0.9185 | 32.18/0.8992 | 32.15/0.9279 |
| ESRT_IMDN_IBMD | ×2 | 545 | 18.40 | 38.03/0.9671 | 33.62/0.9170 | 32.15/0.8990 | 32.16/0.9278 |
| ESRT_RFDN_IBMD | ×2 | 622 | 21.47 | 38.06/0.9673 | 33.66/0.9174 | 32.16/0.8992 | 32.17/0.9277 |
| SRCNN | ×3 | 8 | 52.7 | 32.75/0.9090 | 29.30/0.8215 | 28.41/0.7863 | 26.24/0.7989 |
| VDSR | ×3 | 666 | 612.6 | 33.66/0.9213 | 29.77/0.8314 | 28.82/0.7976 | 27.14/0.8279 |
| DRCN | ×3 | 1774 | 9788.7 | 33.82/0.9226 | 29.76/0.8311 | 28.80/0.7963 | 27.15/0.8276 |
| EDSR | ×3 | 1554 | 160.4 | 34.28/0.9263 | 30.24/0.8405 | 29.06/0.8044 | 28.00/0.8493 |
| FMEN | ×3 | 757 | 77.2 | 34.45/0.9275 | 30.40/0.8435 | 29.17/0.8063 | 28.33/0.8562 |
| ESRT | ×3 | 770 | 126.91 | 34.42/0.9268 | 30.43/0.8433 | 29.15/0.8063 | 28.46/0.8574 |
| IMDN | ×3 | 703 | 71.5 | 34.36 0.9270 | 30.32/0.8417 | 29.09/0.8046 | 28.17/0.8519 |
| RFDN | ×3 | 541 | 55.4 | 34.41/0.9273 | 30.34/0.8420 | 29.09/0.8050 | 28.21/0.8520 |
| BSRN | ×3 | 340 | 33.3 | 34.46/0.9277 | 30.47/0.8449 | 29.18/0.8068 | 28.39/0.8567 |
| RMFDN | ×3 | 224 | 41.0 | 34.30/0.9267 | 30.30/0.8414 | 29.03/0.8035 | 28.00/0.8493 |
| IMDN_ IBMD | ×3 | 194 | 10.91 | 34.35/0.9271 | 30.33/0.8415 | 29.07/0.8045 | 28.13/0.8515 |
| RFDN_ IBMD | ×3 | 178 | 9.33 | 34.42/0.9274 | 30.35/0.819 | 29.10/0.8052 | 28.20/0.8521 |
| ESRT_IMDN_IBMD | ×3 | 915 | 33.84 | 34.43/0.9275 | 30.42/0.8430 | 29.14/0.8061 | 28.44/0.8575 |
| ESRT_RFDN_IBMD | ×3 | 991 | 36.91 | 34.456/0.9276 | 30.44/0.8445 | 29.15/0.8065 | 28.37/0.8563 |
| SRCNN | ×4 | 8 | 52.7 | 30.48/0.8626 | 27.50/0.7513 | 26.90/0.7101 | 24.52/0.7221 |
| VDSR | ×4 | 666 | 612.6 | 31.35/0.8838 | 28.01/0.7674 | 27.29/0.7251 | 25.18/0.7524 |
| DRCN | ×4 | 1774 | 9788.7 | 31.53/0.8854 | 28.02/0.7670 | 27.23/0.7233 | 25.14/0.7510 |
| EDSR | ×4 | 1518 | 114.2 | 31.98/0.8927 | 28.55/0.7805 | 27.54/0.7348 | 25.90/0.7809 |
| FMEN | ×4 | 769 | 44.2 | 32.24/0.8955 | 28.70/0.7839 | 27.63/0.7379 | 26.28/0.7908 |
| ESRT | ×4 | 2992 | 160.31 | 31.23/0.8753 | 28.47/0.7804 | 27.39/0.7336 | 25.96/0.7925 |
| IMDN | ×4 | 715 | 40.9 | 32.21/0.8948 | 28.58/0.7811 | 27.56/0.7353 | 26.04/0.7838 |
| RFDN | ×4 | 550 | 31.6 | 32.24/0.8952 | 28.61/0.7819 | 27.57/0.7360 | 26.11/0.7858 |
| BSRN | ×4 | 352 | 19.4 | 32.35/0.8966 | 28.73/0.7847 | 27.65/0.7387 | 26.27/0.7908 |
| RMFDN | ×4 | 243 | 34.0 | 32.16/0.8940 | 28.51/0.7793 | 27.49/0.7331 | 25.90/0.7798 |
| IMDN_ IBMD | ×4 | 206 | 11.70 | 32.25/0.8944 | 28.56/0.7815 | 27.54/0.7348 | 26.05/0.7836 |
| RFDN_ IBMD | ×4 | 187 | 9.91 | 32.26/0.8955 | 28.60/0.7817 | 27.58/0.7358 | 26.09/0.7855 |
| ESRT_IMDN_IBMD | ×4 | 841 | 67.24 | 32.24/0.8952 | 28.50/0.7809 | 27.51/0.7348 | 26.05/0.7836 |
| ESRT_RFDN_IBMD | ×4 | 917 | 70.33 | 32.23/0.8954 | 28.61/0.7812 | 27.55/0.7356 | 26.10/0.7856 |